\documentclass[twocolumn]{aastex701}
\DeclareUnicodeCharacter{2212}{\ensuremath{-}}

\usepackage{xspace}
\usepackage{multirow}
\usepackage{graphicx}
\usepackage{amsmath}

\newcommand{\fesc}{f_{\rm esc}}
\begin{document}

\title{Does the Escape Fraction Prescription Shape Reionization History and Morphology?}

\author[orcid=0009-0001-3072-0995]{Pulkit Sheoran}
\altaffiliation{Email: \href{mailto:pulkitsheoran022@gmail.com}{pulkitsheoran022@gmail.com}}
\affiliation{School of Physical Sciences, Indian Association for the Cultivation of Science, Kolkata-700032, India}
\email{ug23ps2957@iacs.res.in}

\author[orcid=0000-0001-7072-570X]{Ankita Bera}
\altaffiliation{Email: \href{mailto:ankitabm@umd.edu}{ankitabm@umd.edu}} 
\affiliation{Department of Astronomy, University of Maryland, College Park, MD 20742, USA}
\email{ankitabm@umd.edu}

\author[0000-0002-1050-7572]{Sultan Hassan}
\affiliation{Space Telescope Science Institute, 3700 San Martin Dr, Baltimore, MD 21218}
\email{shassan@stsci.edu}

\begin{abstract}
The escape fraction of ionizing photons, $f_{\rm esc}$, is among the least constrained quantities of the Epoch of Reionization (EoR), and intergalactic opacity prevents its direct measurement at high redshifts. We constrain three power-law parameterizations, $f_{\rm esc}(z)$, $f_{\rm esc}(M_{\rm h})$, and $f_{\rm esc}(z, M_{\rm h})$, against JWST UV luminosity functions and EoR observables (ionizing emissivity, neutral hydrogen fraction, and Thomson optical depth) using Markov Chain Monte Carlo, with priors allowing $f_{\rm esc}$ to increase or decrease with redshift and halo mass. We propagate the best-fit models into the semi-numerical simulation \texttt{SimFast21} and compare their ionization morphology and 21\,cm power spectra with a constant-$f_{\rm esc}$ model. All models reproduce the observations with nearly identical global reionization histories. The data favor $f_{\rm esc}$ decreasing with redshift, from $\approx 0.8$ at $z = 5$ to $\approx 0.35$ at $z = 15$. The halo-mass-only model prefers $f_{\rm esc}$ increasing with halo mass, but this trend is a proxy for redshift evolution. When both dependencies vary jointly, the data prefer a mass-independent $f_{\rm esc}$. At matched global neutral fraction, all four prescriptions yield similar ionization morphology, and the 21\,cm power spectra agree to within $\approx 10\%$. Thus, once calibrated against the same observations, the details of the $f_{\rm esc}$ prescription, set by interstellar medium physics, largely average out in both global observables and the 21\,cm power spectrum. Distinguishing them may require statistics beyond the power spectrum, such as Ly$\alpha$ and 21\,cm forest sightline statistics, or forecasts that exploit their small differences in reionization timing.
\end{abstract}


\section{Introduction}
\label{sec:intro}
The Epoch of Reionization (EoR) marks the last major phase transition of the universe from a neutral to an ionized state \citep[for recent reviews, see][]{Wise2019, Choudhury2022}. The first stars and galaxies, which formed during the Cosmic Dawn, produced copious Lyman continuum (LyC) photons that ionized the neutral hydrogen in the surrounding intergalactic medium (IGM), forming ionized bubbles around them \citep{Barkana2001, Zaldarriaga2004, Dayal2018}. Over time, these bubbles grew and eventually overlapped, marking the completion of reionization. However, most of these ionizing photons could not escape their host galaxies and were instead absorbed by neutral gas and dust in the interstellar medium and circumgalactic medium \citep{Trebitsch2017}. The fraction that does escape into the IGM is referred to as the ionizing photon escape fraction ($f_{\rm esc}$) \citep{Steidel2001, Wyithe2003}. Understanding $f_{\rm esc}$ is critical for establishing both the physical mechanisms and the timeline of reionization. Together with the
ionizing photon production efficiency ($\xi_{\rm ion}$) and the star formation history, it sets the ionizing emissivity, and therefore governs whether reionization proceeded rapidly or gradually and which sources drove it.

Pre-JWST models combined with Planck observations suggest that star-forming galaxies produce enough ionizing photons to reionize the universe by $z \sim 6$. \citet{Finkelstein2019}, for example, argued that a low, halo-mass-dependent escape fraction with a global average of only a few percent ($ \langle f_{\rm esc} \rangle \leq 5 \%$) is sufficient to complete reionization. That conclusion relies on several key assumptions, including star formation extending down to ultra-faint galaxies (a faint-end truncation at $M_{\rm UV} \sim -13$ with photo-suppression quenching star formation below $\log(M_{\rm h}/M_\odot) \sim 9$ in already-ionized regions), an ionizing photon production efficiency that rises modestly toward fainter, higher-redshift dwarfs ($\log \xi_{\rm ion} \sim 25.2$--$25.3$), and a significant but subdominant active galactic nucleus (AGN) contribution at $z \lesssim 7$. However, the JWST observations have revealed an overabundance of UV-luminous galaxies at $z \gtrsim 9$ \citep{Castellano2022, Harikane2022, Finkelstein2022} together with elevated ionizing efficiencies ($\log \xi_{\rm ion} \approx 25.5$--$26.0$) \citep{Endsley2023, Munoz2024}, which imply far more ionizing photons at early times. Taken at face value, these photons would complete reionization too early, in tension with the low CMB Thomson optical depth measured by \citet{Planck2020} and with Lyman-$\alpha$ forest measurements, which place the end of reionization near $z \sim 5.5$ \citep{Becker2015, Bosman2021}, later than the $z \sim 6$ endpoint assumed in earlier models, termed as ``photon budget crisis'' \citep{Munoz2024}. Several resolutions have been proposed, including bursty star formation inflating the observed UV luminosity function \citep{Sun2023}, a shorter ionizing mean free path \citep{Becker2021}, a truncation of the faint-end galaxy population \citep{Munoz2024}, and a revised $\xi_{\rm ion}$ \citep{Simmonds2024}. The possibility we examine here is that the escape fraction of these abundant early galaxies is lower, and/or suppressed toward higher redshift, than low-redshift LyC leaker calibrations imply \citep{flury2022}. 

The IGM is opaque to the ionizing photons at high redshifts, so $\fesc$ cannot be measured directly during this era \citep{Inoue2014}. We therefore rely on models that predict its behavior by constraining it against other available observables \citep[e.g.,][]{Robertson2015, Mitra2015}. Physically, $\fesc$ is set by the interstellar medium (ISM) on scales far below those resolved in reionization simulations. Young stars emit most of their ionizing photons within the first few Myr, while still embedded in dense birth clouds that absorb them \citep{Kimm2014, Ma2015}. Photons escape mainly through low-column-density channels cleared by stellar and supernova feedback, so $\fesc$ peaks only after a delay and varies strongly with the recent star formation history \citep{Kimm2014, Ma2015}.  Processes that extend the ionizing output beyond this delay, such as binary stellar evolution \citep{Ma2016} and runaway massive stars \citep{Kimm2014}, increase the escape fraction, although in the FIRE-2 simulations binaries raise $\langle f_{\rm esc}\rangle$ by only $\approx 25$--35\% \citep{ma2020}. \citet{Howard2017} suggest that the escape fraction is a strongly fluctuating quantity that depends on the mass of the star cluster and the turbulent structure of the gas cloud. \citet{Trebitsch2017}, using a series of high-resolution radiation hydrodynamical simulations, find that $f_{\rm esc}$ in high-redshift galaxies can vary by several orders of magnitude over short timescales, driven primarily by supernova explosions that disrupt dense star-forming clumps and clear low-column-density channels through the halo. 
Since photons leak through such channels, their escape is also direction-dependent, \citet{Schwandt2026} find that anisotropic emission produces smaller, more localized ionized regions than isotropic emission.

Averaged over a galaxy population, the escape fraction depends on the physical properties of the host galaxies \citep{Kimm2014}, which correlate with halo mass ($M_{\rm h}$) \citep{Xu2016}. Simulations of the first galaxies find that feedback clears escape channels more easily in low-mass halos, giving an escape fraction that decreases with halo mass \citep{Wise2014, Paardekooper2015, Xu2016, Kimm2017}, a trend adopted in several reionization models \citep{Park2019, Finkelstein2019}. Because low-mass halos dominate at early times, such models imply a population-averaged escape fraction that increases with redshift, as also assumed in earlier emissivity models \citep{HaardtMadau2012, Kuhlen2012}. However, simulations disagree on the strength and even the sign of the resulting $f_{\rm esc}$--$M_{\rm h}$ trend \citep{ma2020}. We therefore treat any $f_{\rm esc}(M_{\rm h})$ relation as a population-averaged parametrization rather than a description of individual galaxies. While many models adopt a constant $\fesc$ for convenience, this fails to capture the dynamics of galaxy evolution. For example, \citet{Mitra2023} use a data-driven reionization model with a Gaussian Process Regression method to infer a redshift-dependent escape fraction, though that reconstruction relies on pre-JWST data.

To capture the dynamics of the escape fraction, in this paper we consider three simple prescriptions of photon escape fraction that depend on redshift, halo mass, and the both: $f_{\rm esc}(z)$, $f_{\rm esc}(M_{\rm h})$ and $f_{\rm esc}(z, M_{\rm h})$, and compare with a constant $f_{\rm esc}$ baseline model of \citet{Bera2026}. We constrain these models against JWST and EoR observables, namely the UV luminosity functions, the
IGM neutral fraction, and the CMB optical depth, using a Markov Chain Monte Carlo (MCMC) framework to obtain posterior constraints on the model parameters and to trace the evolution of the escape fraction with redshift and/or halo mass. 
We then integrate our MCMC-calibrated $f_{\rm esc}$ prescriptions directly into the semi-numerical simulation \texttt{SimFast21} \citep{Santos2010}.
While full radiative transfer codes such as \texttt{C$^2$-Ray} \citep{Mellema2005} are more physically robust, they are computationally expensive. Other semi-numerical alternatives include excursion-set frameworks \citep[e.g., \texttt{21cmFAST};][]{Mesinger2010}, one-dimensional
radiative transfer schemes \citep[\texttt{GRIZZLY};][]{Ghara2015}, explicitly photon-conserving algorithms \citep[\texttt{SCRIPT};][]{Choudhury2018, Choudhury2022_script}, abundance-matching approaches \citep[\texttt{AMBER};][]{Trac2021}, and models
coupled to semi-analytic galaxy formation \citep[\texttt{ASTRAEUS};][]{Hutter2020}. \texttt{SimFast21} is based on the excursion-set formalism but populates the non-linear density field with resolved halos. This allows us to move beyond globally averaged reionization histories to the full three-dimensional ionization field, quantifying whether,
and to what extent, an evolving escape fraction alters the morphology of ionized bubbles and the resulting 21-cm power spectra at fixed neutral fraction.

This paper is organized as follows. In Section~\ref{sec:model}, we describe the source model, combining the intrinsic ionizing photon production rate
with our three $f_{\rm esc}$ parameterizations. Section~\ref{sec:obs_constraints} describes how the UV luminosity function, ionizing emissivity, neutral
hydrogen fraction, and Thomson optical depth are computed in the model, and lists the observational data used to constrain them. In Section~\ref{sec:computational_methodology}, we present the MCMC framework and our implementation of the $f_{\rm esc}$ models in \texttt{SimFast21}. Section~\ref{sec:results} presents the MCMC constraints on the escape fraction,
followed by the ionization morphology and 21\,cm power spectra from the simulations. We summarize our conclusions and discuss future directions in Section~\ref{sec:conclusion}.
\section{Source model}
\label{sec:model}
In this section we describe the source model that governs the ionizing photon budget, combining a prescription for the intrinsic photon production rate with the redshift- and halo-mass-dependent escape-fraction models.
\subsection{Ionization Rate, \texorpdfstring{$R_{\rm ion}$}{R\_ion}}
To quantify the intrinsic production rate of ionizing photons, we adopt the source model of \citet{Hassan2016}, with the ionization rate, $R_{\rm ion}$ (in units of photons\,s$^{-1}$) as a function of halo mass ($M_{\rm h}$) and redshift ($z$), derived from radiative-transfer simulations \citep{Finlator2015b}. This model has previously been calibrated against EoR observables \citep{Bera2023} and JWST constraints \citep{Bera2026}.
\enlargethispage{\baselineskip}
The ionization rate is given by
\begin{equation}
\label{eq:rion}
    R_{\rm ion} = A \times M_{\rm h} \left( 1 + z \right)^D \left(\frac{M_{\rm h}}{B} \right)^C \exp \left[ - \left(\frac{M_{\rm h}}{B} \right)^{-3} \right]
\end{equation}

The mass dependence in this relation is Schechter-like, a power law at the massive end and an exponential turnover at low masses due to supernova and photoionization feedback, while the redshift dependence is a simple power law.
Here $A$, $B$, $C$ and $D$ are free parameters. The parameter $A$ acts as the overall amplitude factor with units of s$^{-1}\,M_\odot^{-1}$. $B$ is the characteristic quenching mass scale below which photon production is exponentially suppressed by feedback, with units of $M_\odot$. The parameter $C$ sets the slope of the $R_{\rm ion}-M_{\rm h}$ relation, so larger $C$ shifts the ionizing budget toward massive halos. $D$ accounts for the redshift dependence in the ionizing photon production rate.
The rate of ionizing photons that escape a halo is obtained by multiplying $R_{\rm ion}$ with the escape fraction $f_{\rm esc}$.
\subsection{The ionizing photon escape fraction}
\label{subsec:fesc}
The physical dependences of the population-averaged escape fraction remain uncertain, and previous studies have parameterized it as a function of either halo mass or redshift. We consider power-law parameterizations in both, following \citet{Park2019}, who model $f_{\rm esc}$ as a power law in halo mass, and \citet{Faucher2020}, who model it as $f_{\rm esc} \propto (1+z)^{\gamma}$. We define three functional forms: $f_{\rm esc}(z)$, $f_{\rm esc}(M_{\rm h})$, and a combined $f_{\rm esc}(z,M_{\rm h})$.

A redshift-dependent escape fraction has been inferred from data-driven reionization models \citep{Mitra2023} and adopted in models of the UV background \citep{HaardtMadau2012, Faucher2020}. Post-processing of TNG50 likewise finds that the average escape fraction of low-mass galaxies evolves with redshift \citep{Kostyuk2025}. In our first model, $f_{\rm esc}$ depends only on redshift:
\begin{equation}
\label{eq:fesc_z}
    f_{\rm esc}(z) = a \left( \frac{1+z}{6}\right)^{b_{\rm z}},
\end{equation}
where $a$ is the escape fraction at $z=5$ and $b_{\rm z}$ is the redshift power-law index.

A dependence on halo mass is expected because the gas column density around young stars and the efficiency with which feedback clears escape channels both vary with the host. Simulations of the first galaxies find an escape fraction that decreases with halo mass \citep{Wise2014, Paardekooper2015, Xu2016}, while others find a weak \citep{Kimm2014} or non-monotonic \citep{ma2020} dependence. In our second model, $f_{\rm esc}$ is independent of redshift and depends only on halo mass:
\begin{equation}
\label{eq:fesc_m}
    f_{\rm esc}(M_{\rm h}) = a \left( \frac{\log_{10}(M_{\rm h}/M_\odot)}{10}\right)^{b_{\rm m}},
\end{equation}
where $a$ is the escape fraction at the reference mass $M_{\rm h}=10^{10}\,M_\odot$ and $b_{\rm m}$ is the power-law index in logarithmic halo mass.\footnote{We also tested the alternative parameterizations $f_{\rm esc}(M_{\rm h}) = a (M_{\rm h} / 10^{10}\,M_\odot)^{b_{\rm m}}$ and the corresponding combined form $f_{\rm esc}(z, M_{\rm h}) = a ((1+z)/6)^{b_{\rm z}} (M_{\rm h} / 10^{10}\,M_\odot)^{b_{\rm m}}$. These yielded comparable fits to the observables, and we therefore retain the logarithmic formulation for all mass-dependent models.} 

Since low-mass halos increasingly dominate the halo population toward higher redshift, a mass-dependent $f_{\rm esc}$ alone also produces an evolution of the population-averaged escape fraction with redshift. To separate the two dependencies, we consider an escape fraction evolving with both redshift and halo mass:
\begin{equation}
\label{eq:fesc_zm}
    f_{\rm esc}(z,M_{\rm h}) = a \left( \frac{1+z}{6}\right)^{b_{\rm z}}\left( \frac{\log_{10}(M_{\rm h}/M_\odot)}{10}\right)^{b_{\rm m}},
\end{equation}
in which $a$ is the escape fraction at $z=5$ and $M_{\rm h}=10^{10}\,M_\odot$. Parameter combinations that yield $f_{\rm esc} > 1$ are excluded through the prior (Section~\ref{subsec:mcmc}). \citet{Bianco2026} adopt the same functional form with stellar mass, calibrated directly against Lyman-continuum leakage measurements.

\begin{deluxetable*}{llcc}
\tablewidth{0pt}
\tablecaption{Summary of the ionizing photon escape fraction ($f_{\mathrm{esc}}$) models. 
\label{table:model_summary}}
\tablehead{
    \colhead{Model} & 
    \colhead{Description} & 
    \colhead{Functional Form} & 
    \colhead{Free Parameters}
}
\startdata
\rule[-0.8cm]{0pt}{2.4cm}$f_{\mathrm{esc}}(z)$ & \parbox{5.0cm}{Power law in redshift.} & $f_{\mathrm{esc}} = a \left(\frac{1+z}{6}\right)^{b_z}$ & $a, b_z$ \\[0.4cm]
\tableline
\rule[-0.8cm]{0pt}{2.4cm}$f_{\mathrm{esc}}(M_{\rm h})$ & \parbox{5.0cm}{Power law in logarithmic halo mass; no explicit redshift dependence.} & $f_{\mathrm{esc}} = a \left(\frac{\log_{10}(M_{\rm h}/M_\odot)}{10}\right)^{b_m}$ & $a, b_m$ \\[0.4cm]
\tableline
\rule[-0.8cm]{0pt}{2.4cm}$f_{\mathrm{esc}}(z, M_{\rm h})$ & \parbox{5.0cm}{Power law in both redshift and logarithmic halo mass.} & $f_{\mathrm{esc}} = a \left(\frac{1+z}{6}\right)^{b_z} \left(\frac{\log_{10}(M_{\rm h}/M_\odot)}{10}\right)^{b_m}$ & $a, b_z, b_m$ \\[0.2cm]
\tableline
\rule[-0.8cm]{0pt}{2.4cm}Constant $f_{\mathrm{esc}}$ & \parbox{5.0cm}{Same value for all halos and redshifts.} & $f_{\mathrm{esc}} = \mathrm{const.}$ & None (fixed to 0.573)\tablenotemark{a} \\[0.4cm]
\tableline
\enddata
\tablenotetext{a}{Baseline reference model adopted from \citet{Bera2026}.}
\tablecomments{The functional forms represent the theoretical parameterizations of $\fesc$ evaluated in both the MCMC and \texttt{SimFast21} frameworks. The parameters ($A, B, C, D$) are omitted from this summary for clarity but are evaluated simultaneously.}
\end{deluxetable*}

\section{Observational Constraints}
\label{sec:obs_constraints}
To constrain the free parameters of our $f_{\rm esc}$ models, we compare the model predictions against JWST and EoR observables. All data are compiled using the publicly available \texttt{CoReCon} package \citep{Garaldi2023}, following the compilation of \citet{Bera2026}. In the following subsections, we describe
how each observable is computed in the model and list the datasets used in the likelihood (Section~\ref{subsec:mcmc}).
\subsection{The UV luminosity function, UVLF}
\label{subsec:uvlf}
The UV luminosity function, $\phi(M_{\rm AB})$ represents the comoving number density of galaxies per unit absolute magnitude,
\begin{equation}
    \label{eq:uvlf}
    \phi(M_{\rm AB}) = \frac{dn}{dM_{\rm AB}}
    = \frac{dn}{dM_{\rm h}} \left|\frac{dM_{\rm h}}{dM_{\rm AB}}\right|,
\end{equation}
where the second equality assumes a monotonic, one-to-one relation between halo mass and UV magnitude. We compute the halo mass function $dn/d\ln M_{\rm h}$ over $z \in [4, 20]$ with \texttt{Colossus} \citep{Colossus}, adopting the \citet{Sheth1999} model and the friends-of-friends (FoF) mass definition.

To relate $R_{\rm ion}$ to the UV luminosity, we use the ionizing photon production efficiency, $Q_{\rm ion}$, following the sub-solar metallicity-dependent parameterization \citep{2003A&A...397..527S}, given by
\begin{equation}
\label{eq:qion}
    \log_{\rm 10}Q_{\rm ion} (Z) = 0.639(-\log\, Z)^{1/3} + 52.62 - 0.182,
\end{equation}
where Z is the mass-weighted metallicity, which evolves with redshift as \citep{2017ApJ...840...39M}
\begin{equation}
    log\left(Z/Z_{\odot} \right) = 0.153 - 0.074 \, z^{1.34}
\end{equation}

The star formation rate implied by $R_{\rm ion}$ is $R_{\rm ion}/Q_{\rm ion}$, which we convert to an intrinsic UV luminosity (in
erg\,s$^{-1}$\,Hz$^{-1}$) using the conversion factor of \citet{1998ApJ...498..106M}:
\begin{equation}
\label{eq:luv}
    L_{\rm UV} = \frac{R_{\rm ion}}{1.25 \times 10^{-28}\, Q_{\rm ion}}.
\end{equation}

The absolute magnitude is then
\begin{equation}
\label{eq:mab}
    M_{\rm AB} = -2.5 \log_{10}\!\left(\frac{L_{\rm UV}}{\rm erg\,s^{-1}\,Hz^{-1}}\right) + 51.60.
\end{equation}

We use the UVLF measurements of \citealp{2023ApJ...951L...1P,2024ApJ...965..169A,2023MNRAS.523.1009B,2023MNRAS.523.1036B,2024ApJ...965...98C,2023MNRAS.518.6011D,2023MNRAS.520.4554D,2024MNRAS.533.3222D,2024ApJ...969L...2F,2023ApJS..265....5H,2023ApJ...954L..46L,2024MNRAS.527.5004M,2024ApJ...970...31R,2024ApJ...966...74W,2025ApJ...992...63W}.

\subsection{Ionizing Emissivity, \texorpdfstring{$\dot{N}_{\rm ion}$}{N\_ion}}
The ionizing emissivity, $\dot{N}_{\rm ion}$, is the rate at which ionizing photons escape into the IGM per unit comoving volume, and is given by
\begin{equation}
    \label{eq:nion}
    \dot{N}_{\rm ion} = \int_{M_{\rm h,min}}^{\infty}
    R_{\rm esc}(M_{\rm h}, z)\, \frac{dn}{dM_{\rm h}}\, dM_{\rm h},
\end{equation}

where $R_{\rm esc} = f_{\rm esc} \times R_{\rm ion}$ is the escaping ionizing photon rate of a halo of mass $M_{\rm h}$.

We use the emissivity estimates of \citealp{2021MNRAS.508.1853B,2023MNRAS.525.4093G} and \citet{2013MNRAS.436.1023B}.

\subsection{Neutral Hydrogen Fraction, \texorpdfstring{$x_{\rm HI}$}{x\_HI}}
We obtain the reionization history, $x_{\rm HI}(z) = 1 - x_{\rm HII}(z)$, by integrating the volume-averaged ionized hydrogen fraction \citep{1999ApJ...514..648M}:
\begin{equation}
    \label{eq:dxdt}
    \frac{dx_{\rm HII}}{dt} = \frac{\dot{N}_{\rm ion}}{\bar{n}_{\rm H}} - \frac{x_{\rm HII}}{\bar{t}_{\rm rec}} \,.
\end{equation}

The first term is the ratio of the comoving ionizing emissivity to the mean comoving hydrogen number density,
\begin{equation}
    \label{eq:n_h}
    \bar{n}_{\rm H} = \frac{X_{\rm p}\, \Omega_{\rm b}\, \rho_{\rm crit,0}}{m_{\rm H}},
\end{equation}
where $X_{\rm p} = 0.76$ is the primordial hydrogen mass fraction,
$\rho_{\rm crit,0}$ is the present-day critical density, and $m_{\rm H}$ is the hydrogen atom mass. The second term accounts for recombinations, with recombination time
\begin{equation}
\label{eq:trec}
    t_{\rm rec} = \left[\alpha_{\rm A}\, \bar{n}_{\rm H}\, C_{\rm HII}\,
    (1+z)^3\, (1+\chi)\right]^{-1},
\end{equation}
where $\alpha_{\rm A} = 4.2 \times 10^{-13}\,{\rm cm^3\,s^{-1}}$ is the case A recombination coefficient at $T = 10^4\,{\rm K}$ \citep{Kaurov2014}, and $C_{\rm HII}$ is the redshift-dependent clumping factor of \citet{Pawlik2015}, which accounts for the enhancement of the recombination rate by small-scale density inhomogeneities relative to a uniform IGM.

The factor $(1+\chi)$ accounts for electrons from singly ionized helium, assumed to be ionized together with hydrogen, where
\begin{equation}
\label{eq:chi}
    \chi = \frac{Y_{\rm p}}{4(1 - Y_{\rm p})},
\end{equation}
and $Y_{\rm p} = 0.24$ is the primordial helium mass fraction.

We use the neutral fraction constraints derived from a comprehensive suite of observational probes. These include measurements along quasar sightlines like general QSO absorption spectra \citep{2006AJ....132..117F, 2023MNRAS.525.4093G}, QSO damping wings \citep{2018ApJ...864..142D, 2019MNRAS.484.5094G, 2020ApJ...897L..14Y, 2020ApJ...904...26Y, 2024AandA...688L..26S}, and Ly$\alpha$ transmission profiles and forest statistics \citep{2011MNRAS.416L..70B, 2021MNRAS.508.1853B, 2022MNRAS.514...55B}, Ly$\alpha$ damping wings \citep{2017MNRAS.466.4239G, 2024ApJ...971..124U}, Ly$\alpha$ equivalent width (EW) distributions \citep{Mason_2018, 2020ApJ...904..144J, 2024ApJ...967...28N}, Lyman-break galaxies (LBGs) \citep{2019ApJ...878...12H}, and the Lyman-$\alpha$ emitters (LAEs) \citep{2008ApJ...677...12O, 2012ApJ...744...83O}.

\subsection{Thomson optical depth, \texorpdfstring{$\tau$}{tau}}
The CMB photons scatter off the free electrons released during reionization, known as Thomson scattering. The resulting optical depth is
\enlargethispage{\baselineskip}
\begin{equation}
    \label{eq:tau}
    \tau = \int_{\rm 0}^{\infty} dz \, \frac{c(1+z)^2}{H(z)} \, \sigma_{\rm T} \bar{n}_{\rm H} \left[x_{\rm HII}\left(1  + \chi \right)  + \chi \,x_{\rm HeII} \right] ,
\end{equation}
where $c$ is the speed of light, $\sigma_{\rm T}$ is the Thomson cross section, $H(z)$ is the Hubble parameter, and $x_{\rm HeII}$ is the fraction of doubly ionized helium.  

We compare our results to constrain $\tau$ with the \citet{Planck2020} measurement, $\tau = 0.054 \pm 0.007$ ($68\%$ confidence) and $\pm 0.015$ ($95\%$ confidence).

\section{Computational Methodology}
\label{sec:computational_methodology}
To connect our $f_{\rm esc}$ parametrizations to observables, we constrain the model parameters with an MCMC framework and propagate the best-fit models into the semi-numerical code \texttt{SimFast21} to predict the three-dimensional ionization and 21-cm fields.
\subsection{MCMC Implementation}
\label{subsec:mcmc}
We fit each $f_{\rm esc}$ model to the EoR and JWST observables (Section~\ref{sec:obs_constraints}) and sample the posterior with the affine-invariant ensemble sampler \texttt{emcee} \citep{2013PASP..125..306F}. We adopt flat priors on $a \in [0.01, 1]$, $b_z \in [-5, 5]$, $b_m \in [-5, 5]$, $\log_{10} A \in [35, 45]$, $C \in [-10, 10]$, and $D \in [-5, 10]$. The priors on $b_z$ and $b_m$ are symmetric about zero and admit both signs of each dependence, including escape fractions that increase with redshift or decrease with halo mass, as proposed for feedback-regulated low-mass galaxies (Section~\ref{sec:intro}). Following \citet{Bera2023}, we fix $\log_{10}(B/M_\odot) = 7.67$. The sampled parameter vectors are $\theta = \{a, b_z, A, C, D\}$ for $f_{\rm esc}(z)$, $\theta = \{a, b_m, A, C, D\}$ for $f_{\rm esc}(M_{\rm h})$, and $\theta = \{a, b_z, b_m, A, C, D\}$ for $f_{\rm esc}(z, M_{\rm h})$. To keep $f_{\rm esc}$ physical, we assign zero prior probability to any parameter combination that yields $f_{\rm esc} > 1$ within the redshift and halo-mass range of the model.

We compute the log-likelihood for each observable separately and treat them as independent, and hence the total log-likelihood is
\begin{equation}
    \ln \mathcal{L}_{\rm tot}({\theta}) = \ln \mathcal{L}_{\rm x_{\rm HI}} + \ln \mathcal{L}_\tau + \ln \mathcal{L}_{\dot{n}_{\rm ion}} + \ln \mathcal{L}_{\rm UVLF} \,.
    \label{likelihood_total}
\end{equation}

Each observable is modeled as a Gaussian likelihood. For example, the log-likelihood for the neutral hydrogen fraction is
\begin{equation}
    \label{LLxhi}
    {\rm \ln \mathcal{L}_{\rm x_{\rm HI}} (\rm \theta) = - \frac{1}{2} \left( \frac{\rm x_{\rm HI, obs}  - x_{\rm HI, model}} {\rm \sigma_{\rm x_{\rm HI}}} \right)^2   }
\end{equation}

The MCMC framework yields globally averaged quantities, which carry no information about the spatial distribution of ionized and neutral gas. Many distinct distributions can yield the same global averages: for instance, a reionization history driven by many small ionized bubbles can reproduce the same volume-averaged ionized fraction as one driven by a few large bubbles. To test whether $f_{\rm esc}$ models calibrated to the same global observables produce distinguishable ionization morphologies and 21-cm power spectra, we use
\texttt{SimFast21} (Section~\ref{subsec:simfast21}, ~\ref{subsec:21cm}).

\subsection{\textsc{SimFast21}}
\label{subsec:simfast21}
We use the best-fit parameters from the MCMC to generate ionization and 21-cm differential brightness temperature fields with the semi-numerical code \texttt{SimFast21} \citep{Santos2010, Hassan2015, Hassan2016}. \texttt{SimFast21} generates a linear Gaussian random density field, evolves it into the non-linear regime using the \citet{Zel'dovich1970} approximation, and identifies collapsed halos with an excursion-set halo finder applied to the evolved field. The non-linear density field sets both the halo distribution and the density of the surrounding IGM. These are used to compute the ionization rate, which depends on halo mass, and the recombination rate that depends on
the local overdensity, and finally the 21-cm brightness temperature field.

The source model follows \citet{Hassan2016}, who calibrated $R_{\rm ion}$ as a function of halo mass and redshift by combining the high-resolution radiative hydrodynamic simulation of \citet[][6/256-RT]{Finlator2015b} with the larger-volume hydrodynamic simulation of \citet{Dave2013}, and the functional form is given in Eq.~\ref{eq:rion}. The inhomogeneous recombination rate is also calibrated on the 6/256-RT simulation and parametrized as a function of overdensity $\Delta$ and redshift
\citep{Hassan2016},
\begin{equation}
\label{eq:rrec}
    \frac{R_{\rm rec}}{V} = A_{\rm rec}\,(1+z)^{D_{\rm rec}}
    \left[\frac{(\Delta/B_{\rm rec})^{C_{\rm rec}}}
    {1 + (\Delta/B_{\rm rec})^{C_{\rm rec}}}\right]^4,
\end{equation}
with $A_{\rm rec} = 9.85 \times 10^{-24}\,{\rm cm^{-3}\,s^{-1}}$, $B_{\rm rec} = 1.76$, $C_{\rm rec} = 0.82$, and $D_{\rm rec} = 5.07$. In contrast to the global clumping factor used in the MCMC model (Eq.~\ref{eq:trec}), recombinations in \texttt{SimFast21} are therefore computed locally.

\begin{deluxetable*}{lccccc}
\tablewidth{0pt}
\tablecaption{Best-fit values for the three variable $f_{\mathrm{esc}}$ parameterizations, compared against the constant $f_{\mathrm{esc}}$ reference model.}
\label{table:bestfit}
\tablehead{
    \colhead{Parameter} & 
    \colhead{Prior Range} & 
    \colhead{$f_{\mathrm{esc}}(z)$} & 
    \colhead{$f_{\mathrm{esc}}(M_{\rm h})$} & 
    \colhead{$f_{\mathrm{esc}}(z, M_{\rm h})$} & 
    \colhead{Const. $f_{\mathrm{esc}}$\tablenotemark{*}}
}
\startdata
\cutinhead{Escape Fraction Parameters}
$a$ & $[0.01, 1.0]$ & $0.82_{-0.02}^{+0.02}$ & $0.59_{-0.03}^{+0.02}$ & $0.80_{-0.05}^{+0.03}$  &\nodata \\
$b_{\rm z}$ & $[-5.0, 5.0]$ & $-0.90_{-0.08}^{+0.09}$   &\nodata   & $-0.83_{-0.11}^{+0.16}$  &\nodata \\
$b_{\rm m}$ & $[-5.0, 5.0]$ & \nodata    & $1.83_{-0.28}^{+0.23}$  & $0.09_{-0.17}^{+0.32}$  &\nodata \\
\cutinhead{Source Emissivity Model Parameters}
$\log_{10} A$   & $[35.0, 45.0]$ & $35.68_{-0.14}^{+0.16}$    & $35.69_{-0.11}^{+0.19}$   & $35.70_{-0.15}^{+0.18}$ & $35.771_{-0.156}^{+0.153}$  \\
$C$   & $[-10.0, 10.0]$ & $0.56_{-0.03}^{+0.03}$  & $0.58_{-0.03}^{+0.03}$   & $0.56_{-0.03}^{+0.03}$ & $0.612_{-0.023}^{+0.023}$ \\
$D$   & $[-5.0, 10.0]$ & $5.57_{-0.13}^{+0.11}$  & $5.49_{-0.15}^{+0.11}$  & $5.54_{-0.14}^{+0.12}$ & $5.327_{-0.133}^{+0.133}$ \\
\enddata
\tablenotetext{*}{Baseline constant escape fraction model ($f_{\rm{esc}}=0.573$) is adopted from \citet{Bera2026}.}
\tablecomments{The parameter $B$ is fixed to $\log_{10}(B/M_{\odot}) = 7.67$ across all models for likelihood evaluation.}
\end{deluxetable*}
 
We modify the \texttt{SimFast21} code so that the ionization-rate routine evaluates $f_{\rm esc}$ for each halo from Eqs.~\ref{eq:fesc_z}, \ref{eq:fesc_m}, or \ref{eq:fesc_zm}, rather than using a single constant. We set the Sheth-Tormen parameters of the halo finder to $a_{\rm ST} = 0.707$, $p_{\rm ST} = 0.3$, and $A_{\rm ST} = 0.807$ to align them with the mass function used in the MCMC.
The AGN contribution is switched off, consistent with the galaxy-only MCMC
model.
 
To ensure that the volume-averaged ionization history produced by \texttt{SimFast21} matches the history implied by the MCMC best-fit parameters, we introduce a single calibration factor. The escape fractions of Equations~\ref{eq:fesc_z}--\ref{eq:fesc_zm} are multiplied by a constant factor $0.16$ in the simulation. The same factor brings all three models into agreement with their respective MCMC histories, indicating that the offset is a normalization difference between the semi-analytic framework and the simulation rather than a difference in the shape of the evolution.
 
The simulations are run in a box of side length $128\,h^{-1}\,{\rm Mpc}$, with $N_{\rm halo}^3 = 256^3$ cells for the density and halo-finding grid and $N_{\rm smooth}^3 = 256^3$ cells for the smoothed ionization grid.
The minimum halo mass hosting ionizing sources is set to $M_{\rm h,min} = 10^8\,M_\odot$. Outputs are generated over $z \in [4, 20]$ in steps of $\Delta z = 0.2$.

\subsection{21 cm Signal and Power Spectrum}
\label{subsec:21cm}
From the non-linear density, velocity, and ionization fields, \texttt{SimFast21} computes the 21 cm differential brightness temperature relative to the CMB in each cell \citep{Santos2010},
\begin{equation}
\label{eq:dTb}
    \begin{split}
        \delta T_b \approx{}& 23\,x_{\rm HI}\,(1+\delta)
        \left(1 - \frac{T_{\rm CMB}}{T_S}\right)
        \left(\frac{0.7}{h}\right)
        \left(\frac{\Omega_b h^2}{0.02}\right) \\
        &\times \left[\frac{0.15}{\Omega_m h^2}\,\frac{1+z}{10}\right]^{1/2}
        \left[\frac{H(z)}{{\rm d}v_r/{\rm d}r + H(z)}\right]\,{\rm mK},
    \end{split}
\end{equation}
where $x_{\rm HI} = 1 - x_{\rm HII}$ is the neutral fraction, $\delta$ is the non-linear matter overdensity, $T_S$ is the spin temperature, and ${\rm d}v_r/{\rm d}r$ is the gradient of the peculiar velocity along the line of sight, taken along one axis of the box. The velocity gradient is obtained from the non-linear density field using the linear-theory relation and is clipped to $-0.5 \leq ({\rm d}v_r/{\rm d}r)/H \leq 1$ to avoid the divergence at ${\rm d}v_r/{\rm d}r \to -H$.
We assume that the IGM has been heated well above the CMB temperature, $T_S \gg T_{\rm CMB}$, so that the last spin-temperature factor is unity and $\delta T_b$ is set by the neutral-hydrogen density alone. This is the standard assumption during the bulk of reionization. At the earliest stage ($z \simeq 13$) we consider it is an idealization, but because all four models share the same density field and differ only in their $\fesc$ prescriptions, it does not affect the comparison between them.

We characterize the spatial fluctuations of the signal with the spherically averaged dimensionless power spectrum,
\begin{equation}
\label{eq:Delta2}
    \Delta^2(k) = \frac{k^3}{2\pi^2}\,P_{\delta T_b}(k),
\end{equation}
where $P_{\delta T_b}(k)$ is the power spectrum of the $\delta T_b$ field, so that $\Delta^2(k)$ has units of mK$^2$. We compute $P_{\delta T_b}(k)$ with a fast Fourier transform of the $256^3$ grid, after subtracting the mean $\langle \delta T_b \rangle$, and average it over spherical shells of width equal to the fundamental mode, $\Delta k = 2\pi/L \approx 0.049\,h\,{\rm Mpc^{-1}}$.

\section{Results}
\label{sec:results}
In this section we first present the MCMC constraints on the three variable $f_{\rm esc}$ models (Section~\ref{subsec:mcmc_fesc}), and then use the best-fit parameters to drive our \texttt{SimFast21} simulations (Section~\ref{subsec:simfast_results}). Throughout, we compare our results against the constant $f_{\rm esc}$ model of \citet{Bera2026}.
The models are color-coded consistently in all figures: $f_{\rm esc}(z)$ in blue, $f_{\rm esc}(M_{\rm h})$ in green, $f_{\rm esc}(z, M_{\rm h})$ in red, and constant $f_{\rm esc}$ in black, unless otherwise specified.

\subsection{MCMC constraints on the escape fraction}
\label{subsec:mcmc_fesc}

\begin{figure*}[!htbp]
    \centering
    \includegraphics[width=0.98\textwidth]{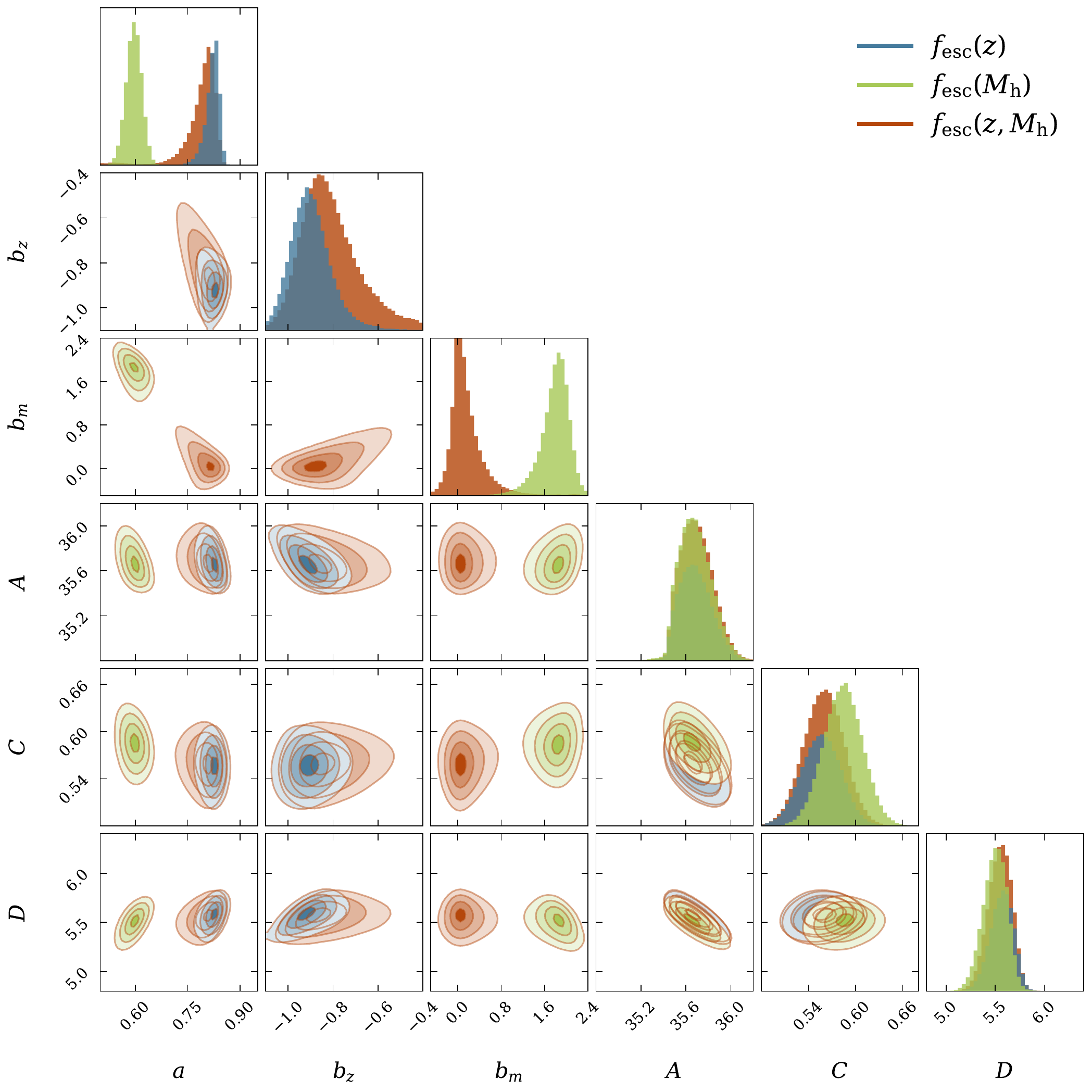}
    \caption{Posterior distributions of the free parameters of the three variable $f_{\rm esc}$ models: $f_{\rm esc}(z)$ (blue), $f_{\rm esc}(M_{\rm h})$ (green), and $f_{\rm esc}(z, M_{\rm h})$ (red). Off-diagonal panels show the 2D marginalized posteriors (68\% and 95\% contours) and diagonal panels the 1D marginalized posteriors. The models are constrained by $x_{\rm HI}$, $\dot{N}_{\rm ion}$, $\tau$ and the UVLF. The best-fit values are summarized in Table~\ref{table:bestfit}. The redshift index $b_z$ is negative in both redshift-dependent models, and the halo-mass index $b_m$ is close to zero once redshift evolution is allowed.}
    \label{fig:corner_plots}
\end{figure*}

The parameterizations in Equations~\ref{eq:fesc_z}, \ref{eq:fesc_m}, and \ref{eq:fesc_zm} are calibrated against the EoR observables $x_{\rm HI}$, $\dot{N}_{\rm ion}$, and $\tau$, together with the JWST UVLF, as described in Section~\ref{sec:obs_constraints}.
Figure~\ref{fig:corner_plots} shows the posterior distributions for the escape-fraction parameters ($a$, $b_z$, $b_m$) and the source-model parameters ($\log_{10} A$, $C$, $D$) of the three parameterizations defined in Section~\ref{subsec:fesc}, and Table~\ref{table:bestfit} lists the parameter values.

\begin{figure}[!htbp]
    \centering
    \includegraphics[width=0.48\textwidth]{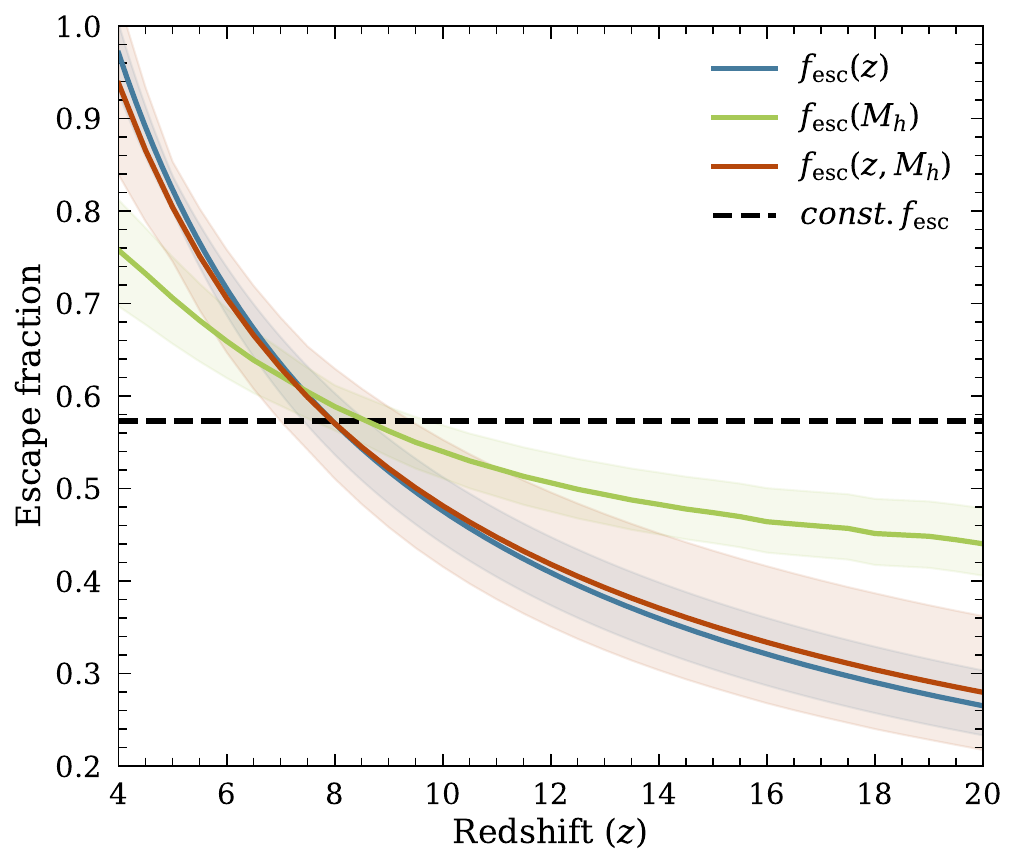}
    \caption{Evolution of the photon escape fraction for our three different models: $f_{\rm esc}(z)$ (blue), $f_{\rm esc}(M_{\rm h})$ (green) and $f_{\rm esc}(z, M_{\rm h})$ (red), together with the constant $f_{\rm esc} = 0.573$ of \citet{Bera2026} (black dashed line). All three variable models favor an escape fraction that decreases with increasing redshift, spanning $f_{\rm esc} \sim 0.3$--$0.9$ for the redshift-dependent models and a narrower range, $f_{\rm esc} \sim 0.45$--$0.75$, for $f_{\rm esc}(M_{\rm h})$. In $f_{\rm esc}(M_{\rm h})$, which has no explicit redshift term, the decline arises from the evolving halo population. The $f_{\rm esc}(M_{\rm h})$ and the $f_{\rm esc}(z, M_{\rm h})$ curves are the weighted average of the escape fraction with the Sheth--Tormen halo mass function multiplied by the intrinsic photon emission rate ($R_{\rm ion}$).}
    \label{fig:fesc_evolution}
\end{figure}

The source-model parameters agree across the three models to within $1\sigma$, and with the constant-$f_{\rm esc}$ values of \citet{Bera2026} to within $\approx1.5\sigma$.
For the escape-fraction parameters, the amplitude $a$ is anticorrelated with the power-law index in all three models: lower values of $a$ are compensated by a higher $b_z$ or $b_m$. This degeneracy direction indicates that the data constrain $f_{\rm esc}$ most tightly at $z > 5$ and $M_{\rm h} > 10^{10}\,M_\odot$ rather than at the reference values. In the redshift-dependent models, $a$ lies close to the upper bound imposed by the physical prior $f_{\rm esc} \le 1$. In the combined model, $b_z$ is constrained away from zero while $b_m$ is consistent with zero. 
A negative $b_z$ and a positive $b_m$ both make the average $f_{\rm esc}$ decline toward higher redshift, the latter because low-mass halos dominate the halo population at early times. The two indices are therefore positively correlated in the combined model (Figure~\ref{fig:corner_plots}): a less negative $b_z$ can be partly compensated by a positive $b_m$. This produces the tails in the $a$ and $b_z$ posteriors toward lower $a$ and less negative $b_z$, which otherwise peak close to those of the $f_{\rm esc}(z)$ model. The trade-off is limited, however, within $1\sigma$, $b_m \lesssim 0.4$ in the combined model, well below the value of $1.83^{+0.23}_{-0.28}$ required when halo mass alone carries the evolution.

Figure~\ref{fig:fesc_evolution} shows the evolution of the escape fraction with redshift. The redshift index $b_z$ is negative in both redshift-dependent models, corresponding to an escape fraction that decreases with increasing redshift.
Within our source model, this implies that galaxies at higher redshift were less transparent to ionizing photons than galaxies at lower redshift. Because the UVLF fixes $R_{\rm ion}/Q_{\rm ion}$ while the EoR observables constrain $f_{\rm esc} R_{\rm ion}$, the quantity constrained by the data is $f_{\rm esc}\,Q_{\rm ion}$. The inferred decline is therefore relative to the evolution of $Q_{\rm ion}$ assumed in Eq.~\ref{eq:qion}, which increases toward higher redshift as the metallicity decreases; an ionizing efficiency that rises less steeply would yield a correspondingly flatter $f_{\rm esc}(z)$.
The decline of $f_{\rm esc}$ with redshift is opposite to the trend found by \citet{Finkelstein2019}, in which faint galaxies dominate the ionizing emissivity and the globally averaged escape fraction increases toward higher redshift, reaching $f_{\rm esc} \approx 0.14$ at $z \sim 15$. \citet{Bianco2026}, calibrating the same functional form as ours also infer the opposite trend, an escape fraction that increases with redshift and decreases with stellar mass. Earlier models of the ionizing background similarly adopted or required an escape fraction that rises with redshift \citep{HaardtMadau2012, Kuhlen2012}. For the $Q_{\rm ion}(z)$ of Eq.~\ref{eq:qion}, our constraints exclude such an increase.
By post-processing the TNG50 simulation, \citet{Kostyuk2025} found that the escape fraction is bimodal, driven by the local gas distribution and metallicity, and that in galaxies with $M_\star < 10^7\,M_\odot$ the average escape fraction decreases with increasing redshift, qualitatively consistent with our redshift-dependent models.

\begin{figure}[!htbp]
    \includegraphics[width=0.48\textwidth]{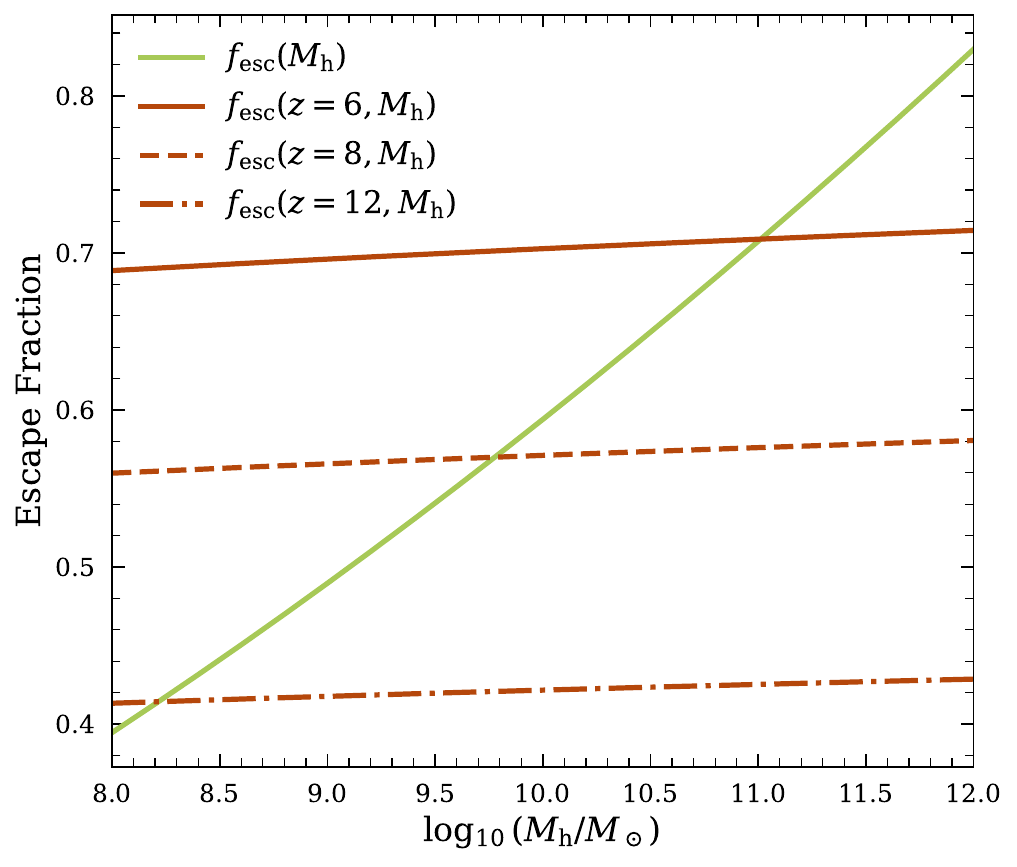}
    \caption{Escape fraction as a function of halo mass for $f_{\rm esc}(M_{\rm h})$ (green) and $f_{\rm esc}(z, M_{\rm h})$ (red), the latter evaluated at $z = 6$ (solid), $8$ (dashed), and $12$ (dot-dashed). The $f_{\rm esc}(M_{\rm h})$ model increases with halo mass, while $f_{\rm esc}(z, M_{\rm h})$ is nearly flat in halo mass and higher at lower redshift, consistent with Figure~\ref{fig:fesc_evolution}.}
    \label{fig:fesc_hm}
\end{figure}

In the $f_{\rm esc}(M_{\rm h})$ model, the posterior favors a positive halo-mass index, $b_{\rm m} = 1.83^{+0.23}_{-0.28}$, despite a prior that admits both signs, so that the escape fraction increases with halo mass (Figure~\ref{fig:fesc_hm}). This contrasts with much of the pre-JWST literature, which favors an escape fraction that decreases with halo mass. In radiation-hydrodynamic simulations of the first galaxies, \citet{Wise2014} find that the luminosity-weighted $f_{\rm esc}$ decreases from $\approx 0.5$ in halos of $10^{6.5}$--$10^{7}\,M_\odot$ to $\approx 0.25$ at $10^{8}\,M_\odot$ and $\approx 0.05$ at $10^{8.5}\,M_\odot$, and \citet{Paardekooper2015} find a similar decline; in both, low-mass galaxies dominate the ionizing photon budget.
\citet{Park2019}, parameterizing both the star formation fraction and the escape fraction as power laws in halo mass, found that the star formation fraction increases with halo mass while the escape fraction decreases, with a power-law index $\alpha_{\rm esc} \approx -0.5$.
Using \texttt{SCRIPT}, \citet{Maity2021} found a declining $f_{\rm esc}$ with a normalization of $0.3$, a slope of $-0.35$, and a reference mass of $10^9\,M_\odot$. Additionally, with the \texttt{DELPHI} framework, \citet{Bremer2023} showed that if photon leakage is regulated by the central gas density, the escape fraction is suppressed in massive, dense halos, and low-mass galaxies ($M_{\rm h} \lesssim 10^9\,M_\odot$) contribute the bulk of the escaping ionizing emissivity. In contrast, the FIRE-2 simulations of \citet{ma2020}, spanning $M_{\rm vir} \sim 10^{8}$--$10^{12}\,M_\odot$ at $z = 5$--12, give a non-monotonic trend: $\langle f_{\rm esc}\rangle$ increases with halo mass up to $\sim 10^{9.5}\,M_\odot$, is nearly constant ($\approx 0.1$--0.2, depending on resolution) up to $\sim 10^{11}\,M_\odot$, and declines at higher masses owing to dust attenuation. Over the halo masses that dominate the ionizing budget in our source model, this trend is closer to our constraints than the declining relations above.

\begin{figure}[!htbp]
    \centering
    \includegraphics[width=0.48\textwidth]{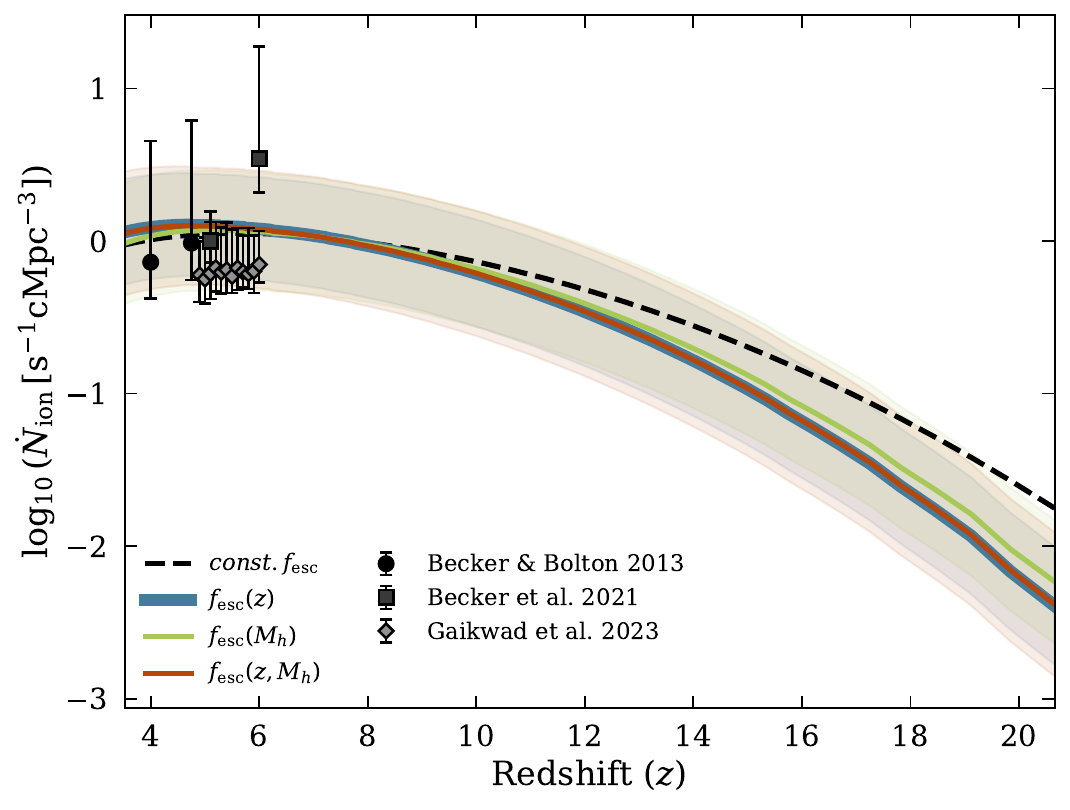}
    \caption{Evolution of the ionizing emissivity $\dot{N}_{\rm ion}$ for the best-fit $f_{\rm esc}(z)$ (blue), $f_{\rm esc}(M_{\rm h})$ (green), and $f_{\rm esc}(z, M_{\rm h})$ (red) models, with 68\% credible intervals. The constant-$f_{\rm esc}$ reference model is shown as a black dashed line. The three variable models show nearly identical evolution, whereas the constant reference model yields a higher an ionizing emissivity at $z \gtrsim 8$. The black markers represent the observational constraints from \citealp{2021MNRAS.508.1853B,2023MNRAS.525.4093G} and \citet{2013MNRAS.436.1023B}.}
    \label{fig:nion_evol}
\end{figure}

\begin{figure}[!htbp]
    \centering
    \includegraphics[width=0.48\textwidth]{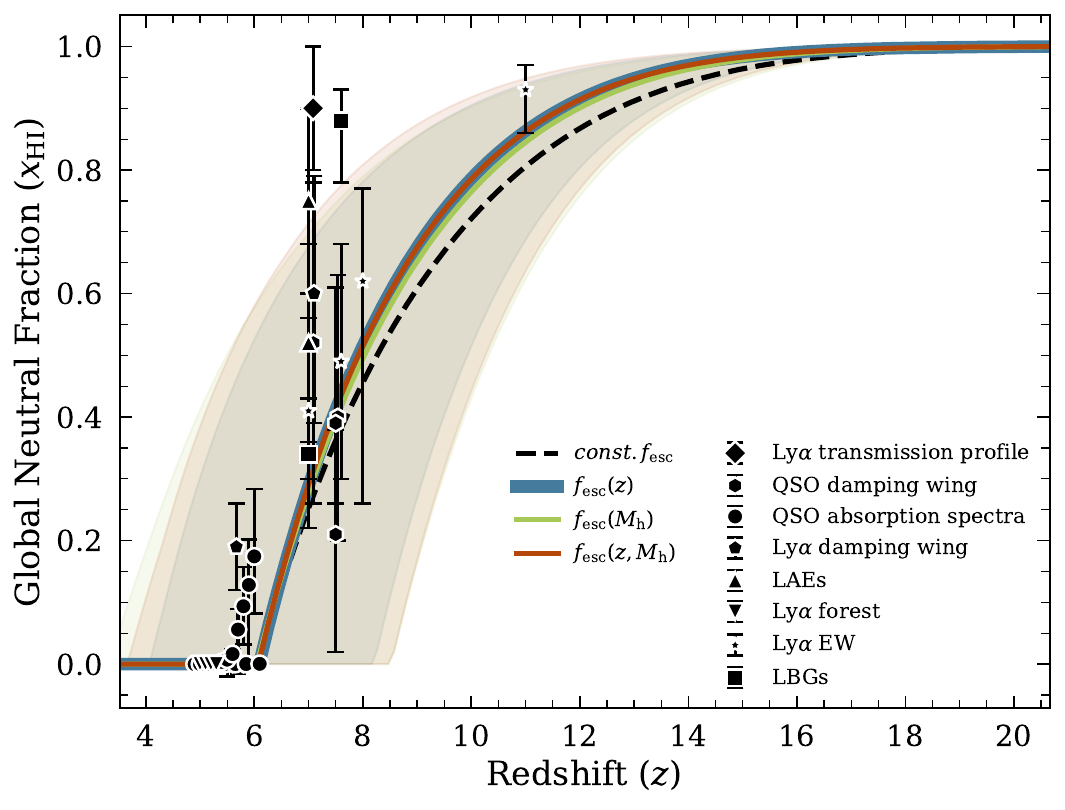}
    \caption{Same as Figure~\ref{fig:nion_evol}, for the volume-averaged neutral hydrogen fraction $x_{\rm HI}$. The observational points include measurements along quasar sightlines like general QSO absorption spectra \citep{2006AJ....132..117F, 2023MNRAS.525.4093G}, QSO damping wings \citep{2018ApJ...864..142D, 2019MNRAS.484.5094G, 2020ApJ...897L..14Y, 2020ApJ...904...26Y, 2024AandA...688L..26S}, and Ly$\alpha$ transmission profiles and forest statistics \citep{2011MNRAS.416L..70B, 2021MNRAS.508.1853B, 2022MNRAS.514...55B}, Ly$\alpha$ damping wings \citep{2017MNRAS.466.4239G, 2024ApJ...971..124U}, Ly$\alpha$ equivalent width (EW) distributions \citep{Mason_2018, 2020ApJ...904..144J, 2024ApJ...967...28N}, Lyman-break galaxies (LBGs) \citep{2019ApJ...878...12H}, and the Lyman-$\alpha$ emitters (LAEs) \citep{2008ApJ...677...12O, 2012ApJ...744...83O}.}
    \label{fig:xhi_evol}
\end{figure}

The positive $b_m$ should not, however, be read as evidence that $f_{\rm esc}$ increases with halo mass. The $f_{\rm esc}(M_{\rm h})$ model carries no explicit redshift dependence, yet its mean declines toward higher redshift (Figure~\ref{fig:fesc_evolution}). This evolution is emergent: at higher redshift the halo population is increasingly dominated by low-mass halos, which have lower $f_{\rm esc}$ in this model. A positive $b_m$ is thus the only way this model can reproduce the decline of $f_{\rm esc}$ with redshift required by the data. When redshift and halo mass are varied simultaneously, $b_m$ becomes consistent with zero ($b_m = 0.09^{+0.32}_{-0.17}$), and the evolution of the escape fraction becomes nearly identical to that of the $f_{\rm esc}(z)$ model (Figure~\ref{fig:fesc_evolution}); correspondingly, the combined model varies only weakly across the halo mass range (Figure~\ref{fig:fesc_hm}). 
We note also that the mass dependence of $f_{\rm esc}$ does not by itself determine which halos dominate the ionizing budget, which depends on the product $f_{\rm esc}(M_{\rm h})\,R_{\rm ion}(M_{\rm h})\,dn/dM_{\rm h}$ integrated over the halo population.

\begin{figure}[!htbp]
    \centering
    \includegraphics[width=0.48\textwidth]{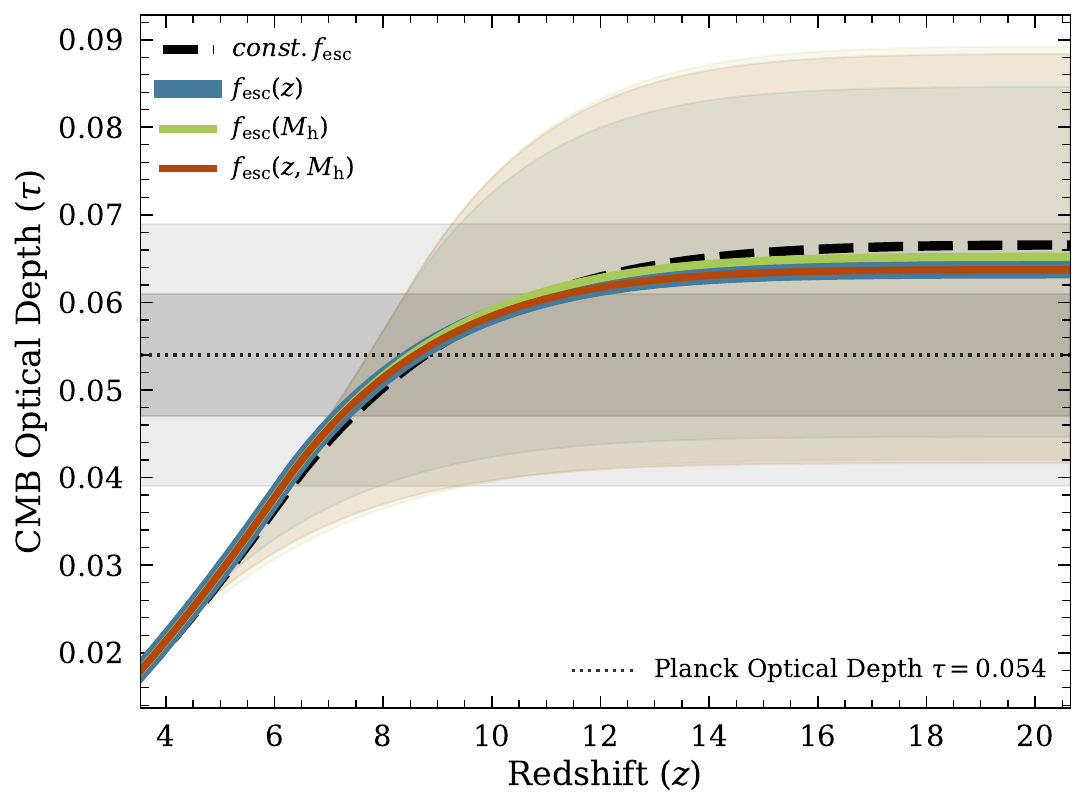}
    \caption{Same as Figure~\ref{fig:nion_evol}, for the cumulative Thomson optical depth $\tau$. The dotted black line represents the \citet{Planck2020} measurement, ($\tau = 0.054$), while the dark and light gray shaded bands indicate the $1\sigma$ ($\pm 0.007$) and $2\sigma$ ($\pm 0.015$) confidence intervals, respectively. All models lie within the 2-$\sigma$ of Planck interval.}
    \label{fig:tau_evol}
\end{figure}

Because all three models reproduce nearly the same $\dot{N}_{\rm ion}(z)$ with nearly the same source-model parameters, the emissivity-weighted $\langle f_{\rm esc}\rangle(z)$ is largely fixed by the data. The amplitudes $a$ are not directly comparable between models, since they are defined at different reference points ($z = 5$ for the redshift-dependent models and $M_{\rm h} = 10^{10}\,M_\odot$ for $f_{\rm esc}(M_{\rm h})$). The residual differences between the weighted curves likely reflect the small shifts in $C$ and $D$ between models (Table~\ref{table:bestfit}).

\begin{figure*}[!htbp]
    \centering
    \includegraphics[height=0.22\textheight]{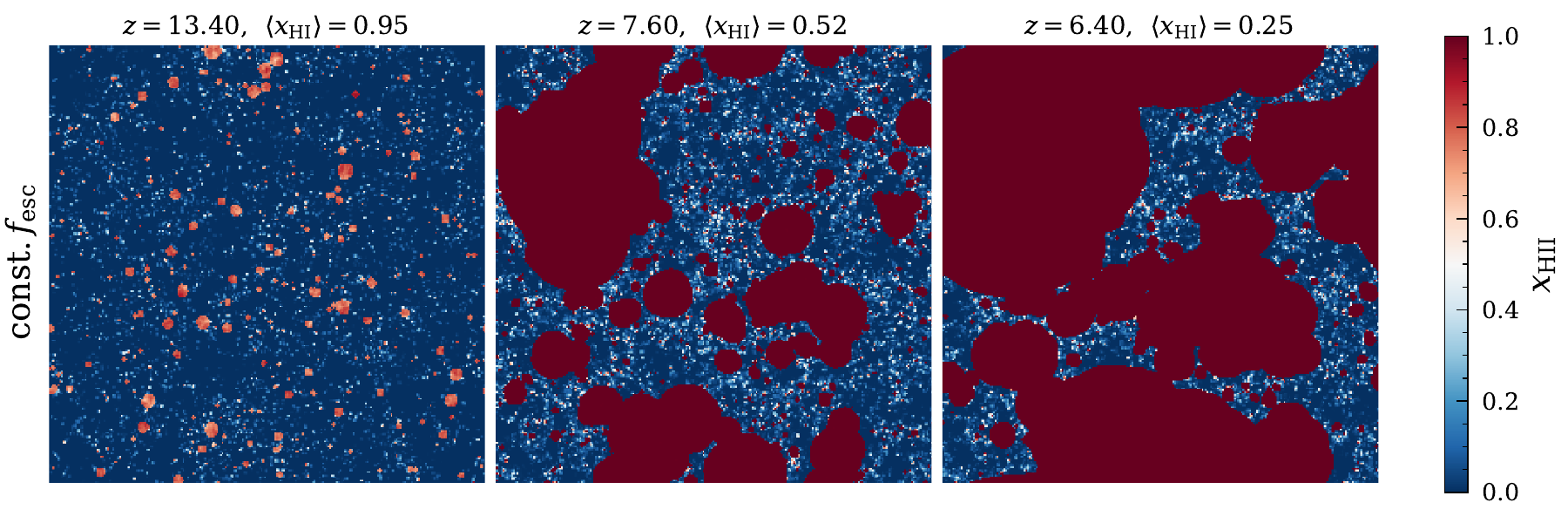}
    \includegraphics[height=0.22\textheight]{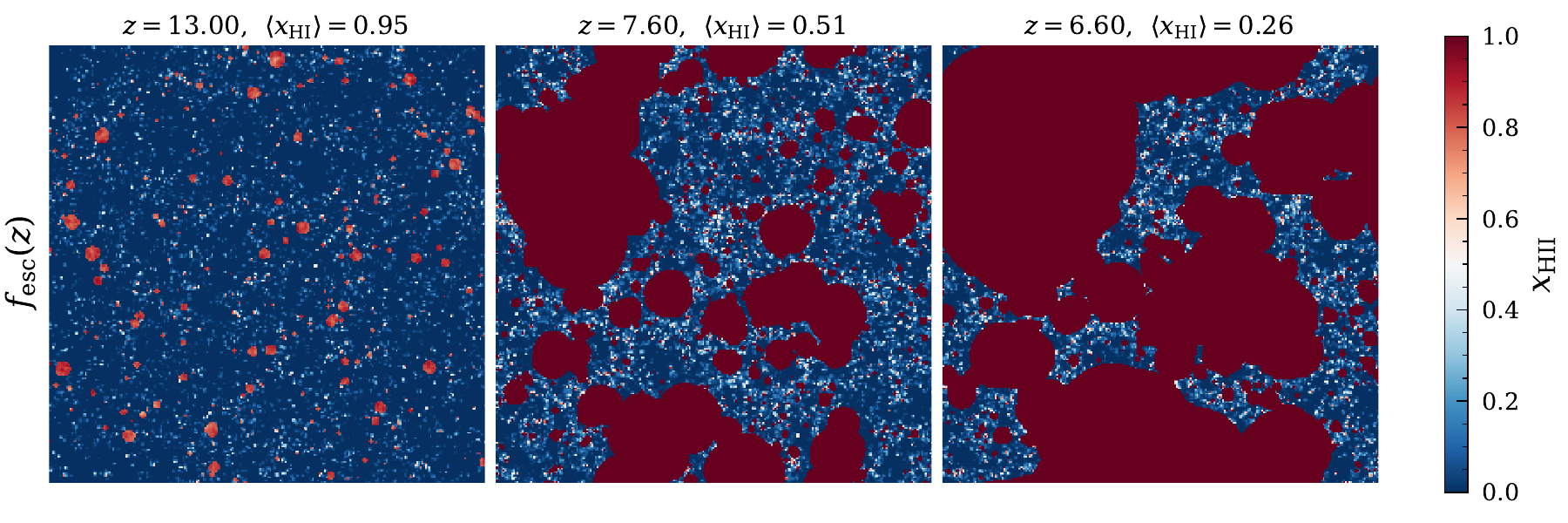}
    \includegraphics[height=0.22\textheight]{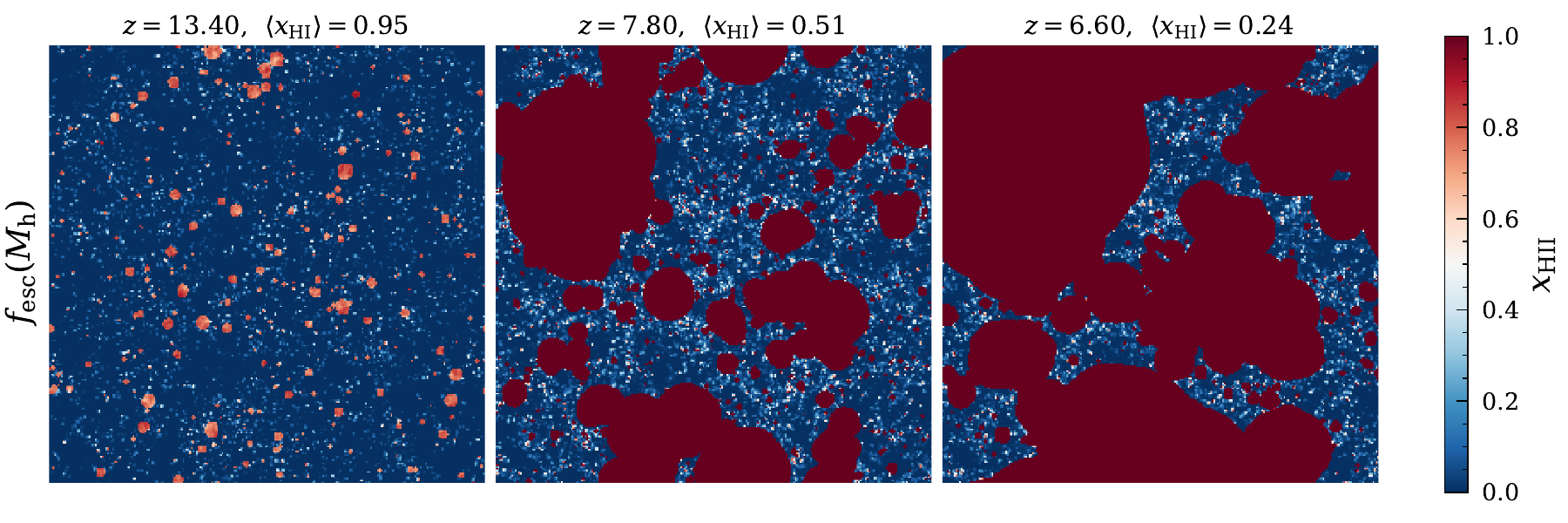}
    \includegraphics[height=0.22\textheight]{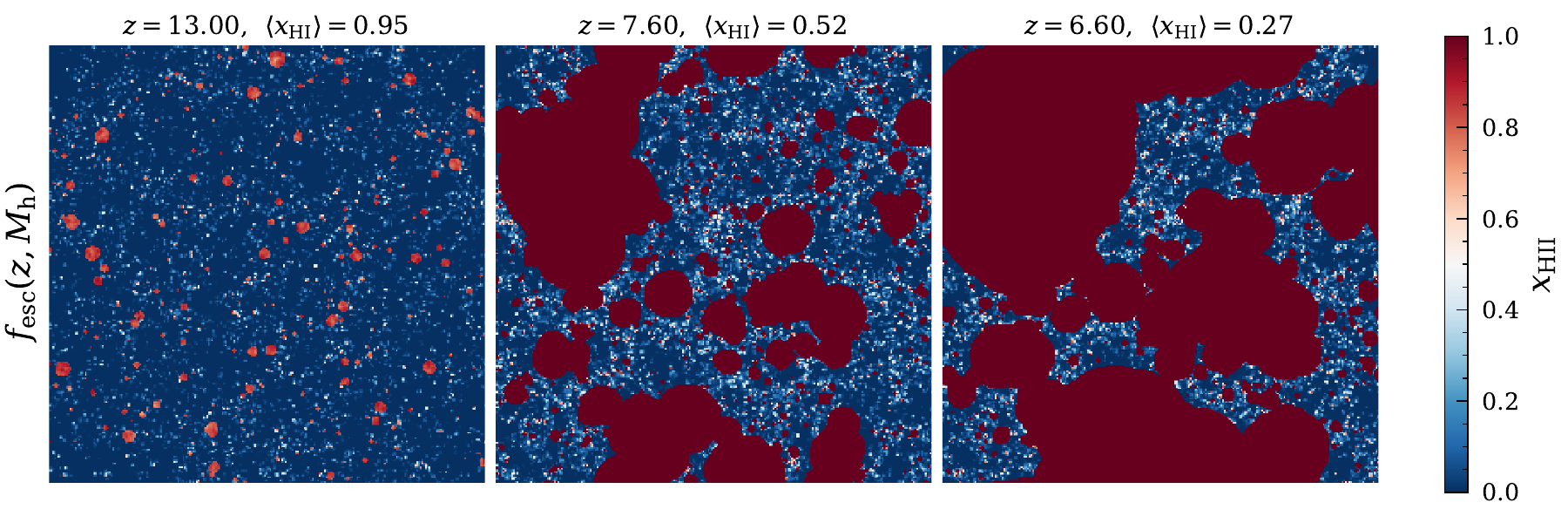}
    \caption{Slices of the ionized fraction $x_{\rm HII}$ through the $(128\,h^{-1}\,{\rm Mpc})^3$ simulation volume, for the constant $f_{\rm esc}$ model (first row), $f_{\rm esc}(z)$ (second row), $f_{\rm esc}(M_{\rm h})$ (third row), and $f_{\rm esc}(z, M_{\rm h})$ (fourth row). Columns correspond to the early ($\langle x_{\rm HI}\rangle\simeq0.95$), intermediate ($\langle x_{\rm HI}\rangle\simeq0.5$), and late ($\langle x_{\rm HI}\rangle\simeq0.25$) stages of reionization. The redshift and global neutral fraction of each snapshot are given above the panel. Because the panels are matched in $\langle x_{\rm HI}\rangle$ rather than in redshift, the corresponding redshifts differ slightly between models. All runs use identical initial conditions.}
    \label{fig:xHII_slices}
\end{figure*}

Despite their different parameterizations, all three models yield a highly consistent evolution of the global EoR observables $\dot{N}_{\rm ion}$, $x_{\rm HI}$, and $\tau$ (Figures~\ref{fig:nion_evol}--\ref{fig:tau_evol}).
The $f_{\rm esc}(z)$ and $f_{\rm esc}(z, M_{\rm h})$ curves are nearly indistinguishable, and all models predict that reionization ends at $z \approx 6$. In contrast, the constant $f_{\rm esc}$ reference model produces a higher ionizing emissivity at early times, so that reionization proceeds faster at high redshift, but it converges to the same completion redshift and yields a very similar optical depth.

\subsection{\texorpdfstring{\textsc{\textup{SimFast21}}}{SimFast21}}
\label{subsec:simfast_results}
%
\subsubsection{Morphology of the ionization field}
\label{subsubsec:xHII_slices}
%
Since the three $f_{\rm esc}$ parameterizations produce nearly identical histories of $\dot{N}_{\rm ion}$,  $x_{\rm HI}$, and $\tau$, the question is whether they remain degenerate at the level of the spatial structure of reionization, which the volume-averaged quantities do not probe. Figure~\ref{fig:xHII_slices} shows slices of the ionized fraction $x_{\rm HII}$ through the simulation volume for the constant $f_{\rm esc}$ model and for each of our three dynamical $f_{\rm esc}$ parameterizations. All four runs share the same initial conditions, so the underlying density field and the locations of ionizing sources are identical by construction, any difference between the panels is therefore attributable to the escape-fraction prescription alone.

The panels in figure~\ref{fig:xHII_slices} are selected at approximately matched global neutral fractions, $\langle x_{\rm HI}\rangle \simeq 0.95$, $0.5$, and $0.25$, rather than at fixed redshift. Comparing at fixed $\langle x_{\rm HI}\rangle$ isolates differences in the morphology of reionization from differences in its timing. The redshifts at which each model reaches a given stage are quoted above each panel.

At all three stages the ionization morphology is closely similar across the four models. The earliest snapshots show small, isolated ionized regions tracing the most massive halos, embedded in an otherwise neutral IGM. By the midpoint of reionization the individual bubbles have grown and begun to overlap into extended ionized regions that follow the filamentary structure of the density field, and by $\langle x_{\rm HI}\rangle \simeq 0.25$ the ionized regions overlap, leaving the remaining neutral gas confined to under-dense regions between them. This is the standard inside-out topology expected for galaxy-driven reionization, and it is reproduced by every escape-fraction prescription we consider. In particular, the $f_{\rm esc}(z)$ and $f_{\rm esc}(z, M_{\rm h})$ slices are visually indistinguishable from one another at all three stages, as expected given that the combined model has a halo-mass index consistent with zero (Section~\ref{subsec:mcmc_fesc}).

The models differ only in the redshift at which a given stage is reached, and these differences are small. The mass-dependent model is the most advanced at fixed redshift, reaching $\langle x_{\rm HI}\rangle \simeq 0.5$ at $z=7.8$ and $\langle x_{\rm HI}\rangle \simeq 0.24$ at $z=6.6$, while the constant $f_{\rm esc}$ model reaches the same stages at $z=7.6$ and $z=6.4$ respectively. The two redshift-dependent models begin slightly later than the others, reaching $\langle x_{\rm HI}\rangle \simeq 0.95$ at $z=13.0$ rather than $z=13.4$, consistent with their negative redshift index suppressing $f_{\rm esc}$ at early times. All of these offsets are $\Delta z \lesssim 0.4$, comparable to a small multiple of the output spacing $\Delta z = 0.2$, and the corresponding continuous ionization histories agree to well within the observational uncertainties (Figure~\ref{fig:xhi_evol}).
 
\subsubsection{The \texorpdfstring{21\,cm}{21 cm} power spectrum}
\label{subsubsec:pk}
%
The ionization slices of Figure~\ref{fig:xHII_slices} suggest that the four escape-fraction prescriptions produce similar spatial structure. We therefore compute the spherically averaged dimensionless power spectrum of the 21\,cm brightness temperature using Eq.~\ref{eq:Delta2} for each model at the same three stages of reionization used in Figure~\ref{fig:xHII_slices}.

As with the slices, the comparison is made at matched global neutral fraction rather than at fixed redshift. This distinction is essential. Evaluated at a common redshift the models differ in $\Delta^2(k)$ by a factor of several, but this reflects the fact that they reach a given stage of reionization at slightly different times: the mean brightness temperature scales as $\langle \delta T_{\rm b}\rangle \propto \langle x_{\rm HI}\rangle$ to leading order, so the power scales as its square, and a modest offset in ionization history produces a large offset in amplitude. Matching in $\langle x_{\rm HI}\rangle$ removes this timing effect and isolates differences in the spatial distribution of the neutral gas.

We quantify these differences in Table~\ref{table:pk_deviation}, which lists,
for each model and stage, the median and maximum fractional deviation of $\Delta^2(k)$ from the constant-$f_{\rm esc}$ model, $|\Delta^2_{\rm model}(k)/\Delta^2_{\rm const}(k) - 1|$, over $k > 0.2\,h\,{\rm Mpc}^{-1}$. The $\langle x_{\rm HI}\rangle \simeq 0.95$ stage serves as a test: the
ionization field is negligible at this point, so any deviation reflects the snapshot selection rather than differences in ionization morphology.

\begin{figure*}[!htbp]
    \centering
    \includegraphics[width=0.98\textwidth]{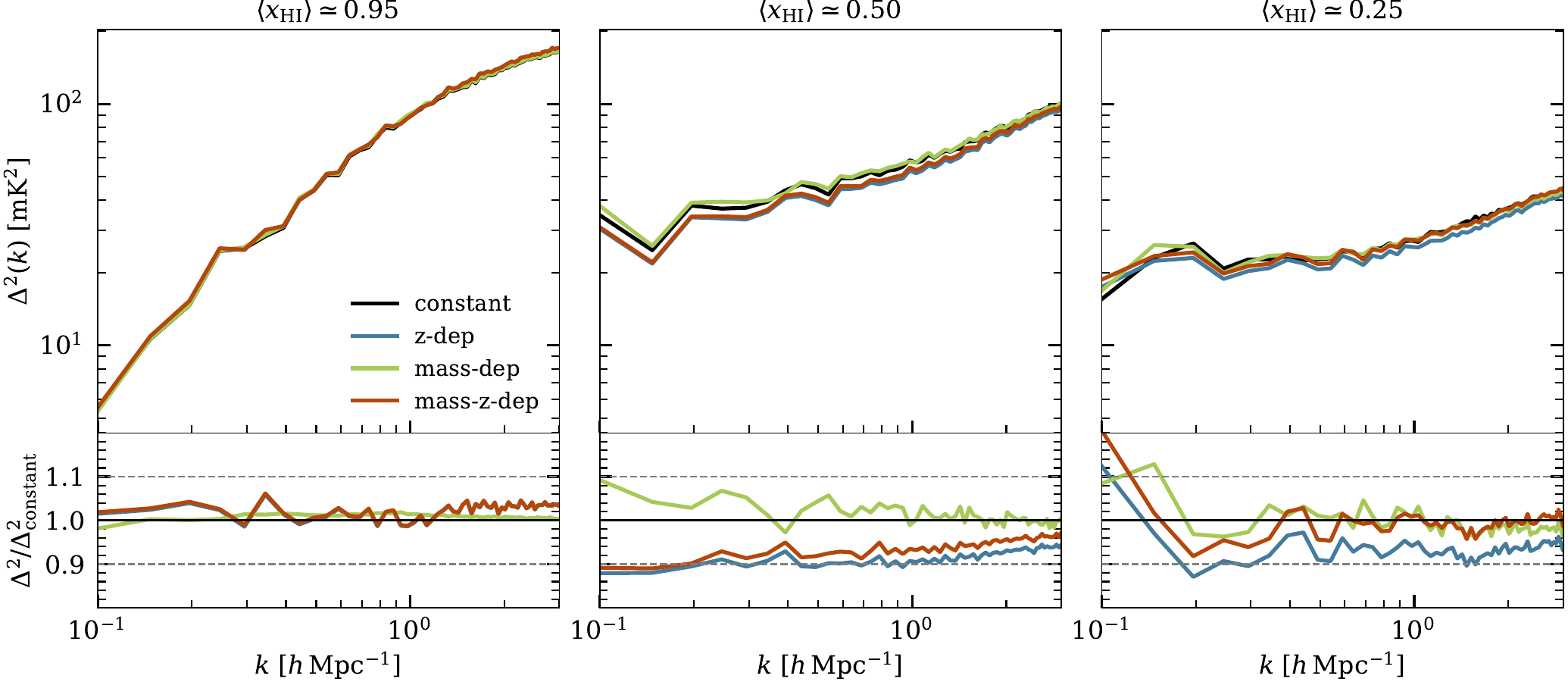}
    \caption{Dimensionless 21\,cm power spectrum $\Delta^2(k)$ for the constant $f_{\rm esc}$ model (black), $f_{\rm esc}(z)$ (blue), $f_{\rm esc}(M_{\rm h})$ (green), and $f_{\rm esc}(z, M_{\rm h})$ (red), evaluated at matched global neutral fractions $\langle x_{\rm HI}\rangle \simeq 0.95$, $0.50$, and $0.25$ (left to right). All models share the same initial conditions. Modes above $k_{\rm Nyq}/2$ are excluded to avoid aliasing.}
    \label{fig:pk_matched}
\end{figure*}

Figure~\ref{fig:pk_matched} and Table~\ref{table:pk_deviation} show that at the earliest stage, $\langle x_{\rm HI}\rangle \simeq 0.95$, the models agree to within $6\%$ for $k > 0.2\,h\,{\rm Mpc}^{-1}$, with a median deviation of $3\%$ from the constant $f_{\rm esc}$ case. The two redshift-dependent models, $f_{\rm esc}(z)$ and $f_{\rm esc}(z, M_{\rm h})$, track one another closely, consistent with the near-zero halo-mass index recovered for the combined model (Section~\ref{subsec:mcmc_fesc}) and with their indistinguishable ionization slices. These are also the two models that reach $\langle x_{\rm HI}\rangle \simeq 0.95$ at $z = 13.0$ rather than $z = 13.4$. With only a few percent of the volume ionized at this point, the brightness-temperature field traces the density field almost directly, so that $\Delta^2 \propto \langle \delta T_{\rm b}\rangle^2\,\Delta^2_\delta \propto (1+z)\,D^2(z) \propto (1+z)^{-1}$ in the matter-dominated era, for $T_{\rm S} \gg T_{\rm CMB}$. The redshift offset alone therefore accounts for a $\approx 3\%$ difference in amplitude, matching the median deviation of these two models. The $f_{\rm esc}(M_{\rm h})$ model, selected at the same
redshift as the constant model, deviates by only $0.8\%$ (median) and $1.9\%$
(maximum), which sets the residual from imperfect matching in $\langle x_{\rm HI}\rangle$ at $\lesssim 1\%$.

At the midpoint and late stages the differences grow only modestly. At $\langle x_{\rm HI}\rangle \simeq 0.50$ the models depart from the constant $f_{\rm esc}$ reference by at most $10.7\%$, with a median deviation of $7.7\%$ for the model that differs most. At $\langle x_{\rm HI}\rangle \simeq 0.25$ the corresponding figures are $10.5\%$ and $6.6\%$. At the midpoint the largest deviations come from the two
redshift-dependent models, and at the late stage, from $f_{\rm esc}(z)$ alone, while the combined model lies within $1.2\%$ (median) of the constant
reference. The $f_{\rm esc}(M_{\rm h})$ model remains close to the constant
reference at every stage, with median deviations below $2\%$. These
differences exceed the $\lesssim 1\%$ matching residual and represent a
genuine but small effect of the escape-fraction prescription on the 21\,cm
power spectrum, although part of the deviation of the redshift-dependent
models may again reflect redshift offsets between the matched snapshots.
 
\begin{deluxetable}{lccc}
\tablewidth{\linewidth}
\tablecaption{Median (maximum) fractional deviation of $\Delta^2(k)$ from the constant $f_{\rm esc}$ model, evaluated at matched neutral fraction over $k > 0.2\,h\,{\rm Mpc}^{-1}$. \label{table:pk_deviation}}
\tablehead{
    \colhead{Model} &
    \colhead{$\langle x_{\rm HI}\rangle \simeq 0.95$} &
    \colhead{$\simeq 0.50$} &
    \colhead{$\simeq 0.25$}
}
\startdata
$f_{\rm esc}(z)$          & 2.9\% (5.9\%) & 7.7\% (10.7\%) & 6.6\% (10.5\%) \\
$f_{\rm esc}(M_{\rm h})$  & 0.8\% (1.9\%) & 1.0\% (6.8\%)  & 1.9\% (4.6\%)  \\
$f_{\rm esc}(z,M_{\rm h})$& 2.9\% (6.2\%) & 5.1\% (8.8\%)  & 1.2\% (6.1\%)  \\
\enddata
\tablecomments{The $\langle x_{\rm HI}\rangle \simeq 0.95$ column serves as a test: at this stage the ionization field is negligible and the quoted values reflect the residual mismatch in matching and finite mode sampling.}
\end{deluxetable}

A further feature is common to all models. At $\langle x_{\rm HI}\rangle \simeq 0.95$ the power spectrum rises steeply and monotonically with $k$, following the density field. At the two later stages the low-$k$ end flattens, reflecting the growth of ionized regions, which suppresses power on scales comparable to and larger than the typical bubble size. The scale at which this flattening occurs is similar in all four models, indicating that the characteristic bubble size at fixed $\langle x_{\rm HI}\rangle$ is not strongly sensitive to how the escape fraction is parameterized.

Taken together with the ionization slices, this establishes the central result of our simulations. Once each parameterization is calibrated against the same EoR and UVLF observations, the resulting 21\,cm signal agrees to within roughly $10\%$ at matched stages of reionization, compared with a $\lesssim 1\%$ residual from the matching procedure itself. The escape-fraction prescriptions are therefore strongly, though not perfectly, degenerate in the spatial statistics of the 21\,cm field as well as in the globally averaged observables of Section~\ref{subsec:mcmc_fesc}. This agrees with \citet{Hassan2022}, who found that replacing one-to-one source models with sources drawn from the \textsc{Simba} hydrodynamic simulation, including scatter in the ionizing emissivity and escape fraction, leaves the shape of the 21\,cm power spectrum unchanged and alters its amplitude by $\lesssim 5$--$10\%$ throughout reionization. In both cases the large-scale structure of reionization at fixed $\langle x_{\rm HI}\rangle$ is insensitive to the details of the source model, while the timing of reionization retains a residual sensitivity, here $\Delta z \lesssim 0.4$ between prescriptions. Similarly, \citet{Bianco2026} find that stochastic and deterministic source models with matched total ionizing photon production yield 21\,cm power spectra within $\approx 8\%$ and nearly identical bubble size distributions, whereas models whose emissivity is distributed differently across halos yield distinct 21\,cm statistics at matched $\langle x_{\rm HI}\rangle$. The residual differences in our models are well below current upper limits from HERA \citep{2023ApJ...945..124H} and LOFAR \citep{2025A&A...698A.186M}. Whether SKA-Low \citep{2026aask.bookE.134A, 2026aask.bookE.136B} could distinguish the prescriptions requires a forecast including thermal noise and foreground contamination, which we defer to future work.
 
\section{Conclusions \& Discussion}
\label{sec:conclusion}
JWST observations have revealed an abundance of UV-luminous galaxies at $z \gtrsim 9$ which, combined with the source models of the pre-JWST era, would overproduce ionizing photons and complete reionization too early. In this work we tested whether a variable ionizing photon escape fraction can accommodate these observations. We parameterized $f_{\rm esc}$ as a function of redshift, halo mass, and both, with priors that allow either sign of each dependence, constrained each parameterization against JWST UV luminosity functions and EoR observables ($x_{\rm HI}$, $\dot{N}_{\rm ion}$, $\tau$) with an MCMC framework, and propagated the best-fit models into the semi-numerical simulation \texttt{SimFast21} to compare the resulting ionization morphology and 21\,cm power spectra with those of a constant-$f_{\rm esc}$ reference model. Our main findings are as follows.

\begin{enumerate}
    \item The data require an escape fraction that decreases with increasing redshift, from $\approx 0.8$ at $z = 5$ to $\approx 0.35$ at $z = 15$. This behavior is common to all three variable models; in the $f_{\rm esc}(M_{\rm h})$ model it emerges through the evolving halo population rather than through an explicit redshift term. Although the priors admit both signs, an escape fraction that increases with redshift, as adopted in earlier reionization models \citep{HaardtMadau2012, Kuhlen2012, Finkelstein2019}, is excluded (see Figure~\ref{fig:fesc_evolution} and Table~\ref{table:bestfit}). Because the UVLF fixes the intrinsic ionizing photon production while the EoR observables constrain the escaping emissivity, this inference is conditional on the assumed ionizing photon production efficiency.

    \item The halo-mass-only model prefers a positive halo-mass index, but this should not be read as evidence that massive halos leak ionizing photons more efficiently: it is the only way this model can reproduce the decline of $f_{\rm esc}$ with redshift, because low-mass halos increasingly dominate the halo population at early times, and the corresponding local slope is shallow ($d\ln f_{\rm esc}/d\ln M_{\rm h} \approx 0.08$ at $10^{10}\,M_\odot$). When redshift and halo mass are varied simultaneously, the halo-mass index is consistent with zero, and the data attribute the evolution of the global escape fraction to redshift rather than to halo mass. An escape fraction that decreases steeply with halo mass, as found in simulations of feedback-regulated low-mass galaxies \citep{Wise2014, Paardekooper2015, Xu2016}, is disfavored, although a mild decrease is not excluded (see Figure~\ref{fig:corner_plots}).

    \item All three variable models predict a nearly identical evolution of the global EoR observables, with reionization ending at $z \approx 6$. The constant-$f_{\rm esc}$ model produces a higher ionizing emissivity at early times but converges to the same completion redshift and a very similar optical depth (see Figures~\ref{fig:nion_evol}--\ref{fig:tau_evol}).

    \item At matched global neutral fraction, the ionization morphology produced by \texttt{SimFast21} is closely similar for all four prescriptions, and the models reach a given stage of reionization within $\Delta z \lesssim 0.4$ of one another (see Figure~\ref{fig:xHII_slices}).

    \item The 21\,cm power spectra agree to within $\approx 10\%$ at matched stages, compared with a $\lesssim 1\%$ residual from the matching procedure, and the $f_{\rm esc}(M_{\rm h})$ model remains closest to the constant reference, with median deviations below $2\%$ (see Figure~\ref{fig:pk_matched} and Table~\ref{table:pk_deviation}). This level of agreement is consistent with that found by \citet{Hassan2022} when replacing simple source models with sources from the \textsc{Simba} simulation.

    \item Once calibrated against the same EoR and UVLF observations, the escape-fraction prescriptions are therefore strongly, though not perfectly, degenerate in both the globally averaged observables and the spatial statistics of the 21\,cm field. Although $f_{\rm esc}$ is set by small-scale physics in the interstellar medium, these large-scale observables are sensitive mainly to its population-averaged value, which the calibration fixes to nearly the same evolution in all prescriptions, and only weakly to how $f_{\rm esc}$ is distributed among the halos resolved here. These observables alone therefore cannot distinguish a constant from a variable escape fraction, nor determine whether the evolution of $f_{\rm esc}$ is driven by halo mass or by redshift.
\end{enumerate}

Several extensions would sharpen these conclusions. Simulations at higher resolution, resolving the low-mass halos in which other studies find the strongest dependence of $f_{\rm esc}$ on halo mass, would test whether the degeneracy persists when these sources are included.
Statistics beyond the power spectrum, such as the bubble size distribution, the bispectrum, and morphological measures, may be more sensitive to differences in ionization topology \citep{Hutter2020, 2026aask.bookE.134A}, as may sightline statistics such as opacity fluctuations in the Ly$\alpha$ forest \citep{Kulkarni2019} and the 21\,cm forest toward radio-loud sources \citep{Furlanetto2006}. Because 21\,cm experiments measure the signal at fixed redshift rather than fixed neutral fraction, forecasts for SKA-Low that include the timing differences between models are needed to assess observability \citep{2026aask.bookE.136B}. Finally, linking $f_{\rm esc}$ to observable galaxy properties, for example through the empirical relation with the UV continuum slope $\beta$ \citep{2022MNRAS.517.5104C}, with its coefficients treated as free parameters, or freeing the ionizing photon production efficiency, would help break the degeneracy between $f_{\rm esc}$ and $\xi_{\rm ion}$ that limits the interpretation of the absolute escape fractions.

\begin{acknowledgments}
PS acknowledges the financial support from the Indian Association for the Cultivation of Science.
AB acknowledges the financial support from the CTC fellowship at the University of Maryland. This work used Bridges-2 at the Pittsburgh
Supercomputing Center through allocation PHY250314 from the Advanced
Cyberinfrastructure Coordination Ecosystem: Services \& Support (ACCESS) program, which is supported by U.S. National Science Foundation grants.
\end{acknowledgments}

\software{
Astropy \citep{2013A&A...558A..33A}, Colossus \citep{2018ApJS..239...35D}, CoReCon \citep{Garaldi2023}, emcee \citep{2013PASP..125..306F}, Matplotlib \citep{2007CSE.....9...90H}, Numpy \citep{2020Natur.585..357H}, SciPy \citep{2020NatMe..17..261V}, 
}

\bibliography{citations}{}
\bibliographystyle{aasjournalv7}

\end{document}